\documentclass[twocolumn]{spie}  

\usepackage{physics}
\usepackage{xcolor}
\usepackage{graphicx}
\usepackage{dcolumn}
\usepackage{bm}
\usepackage{amsmath,amsfonts,amssymb}
\usepackage[colorlinks=true, allcolors=blue]{hyperref}

\title{Effective super-bandwidth in laser pulses}

\author[1,*]{Enrique G. Neyra}
\author[1,2]{Demian A. Biasetti}
\author[1]{Pablo Vaveliuk}
\author[1]{Gustavo A. Torchia}
\author[1,2]{Fabian Videla}
\author[3,4]{Marcelo F. Ciappina}
\author[2,5]{Lorena Rebón}
\affil[1]{Centro de Investigaciones \'Opticas (CICBA-CONICET-UNLP), Cno. Parque Centenario y 506, P.O. Box 3, 1897 Gonnet, Argentina}
\affil[2]{Departamento de Ciencias B\'asicas, Facultad de Ingenier\'ia UNLP, 1 y 47 La Plata, Argentina}
\affil[3]{Physics Program, Guangdong Technion--Israel Institute of Technology, Shantou, Guangdong 515063, China}
\affil[4]{Technion -- Israel Institute of Technology, Haifa, 32000, Israel}

\affil[5]{Instituto de F\'isica de La Plata, CONICET-UNLP, Diagonal 113 entre 63 y 64, La Plata (1900) - Buenos Aires - Argentina}
\affil[*]{enriquen@ciop.unlp.edu.ar}

\begin{document} 
\maketitle

\textbf{We present here a theoretical analysis of the interaction between an ideal two-level quantum system and a super-oscillatory pulse, like the one proposed and successfully synthesized in Ref.~\cite{neyra2021tailoring,arXivsubfourier}. As a prominent feature, these pulses present a high efficiency of the central super-oscillatory region in relation to the unavoidable side-lobes.
Besides, our study shows an increase of the effective bandwidth of the pulse, in the super-oscillatory region, and not only the appearance of a local frequency higher than its highest Fourier-frequency component, as in the usual description of the phenomenon of super-oscillations.
Beyond introducing the concept of effective super-bandwidth, the presented results could be relevant for experimental applications and opening new perspectives for laser-matter interaction.}


\hrule
The super-oscillatory phenomenon~\cite{aharonov2015mathematics}, which intrinsically has an interferometric origin, could be explained as a consequence of the counterintuitive property of a band-limited function that, surprisingly, can vary locally faster than its highest frequency component. It is said that such a behavior locally breaks 
the limit imposed by the Fourier transform (FT), which is restricted by the duration-bandwidth product. Although it is not formally accepted within Fourier's theory since FT is defined in the entire time domain, this paradoxical fact  
is not a mere mathematical curiosity of those kind of functions but it has physical implications with applications to the design of super-directive antennas~\cite{Woodward1948,Wong2011}, quantum weak measurements~\cite{aharonov1988_WM,zhu2019measuring}, and super-resolution microscopy \cite{Berry_2006,rogers2012super}, among others~\cite{Berry_2019}.

Particularly, in the time domain, numerical simulations predict the transmission of a super-oscillating optical signal at the absorption resonance of a dielectric medium~\cite{Eliezer2014}. While the absorption only acts
on the Fourier components, the super-oscillation shows quasi-periodical revivals. 
In the same direction, the transmission of super-oscillating electrical signals at frequencies above the effective band limit of a commercial low-pass filter has been proved in Ref.~\cite{zarkovsky2020transmission}.
This supports the concept of temporal super-resolution introduced in Ref.~\cite{Boyko2005} by analogy to the optical super-resolution imaging technique. 
Hence, a proper shape and spectral control of an ultra-short laser pulse can result in pulses shorter than a transformed-limited Gaussian pulse with the same spectral width. For instance, in Ref.~\cite{Eliezer2017} a super-oscillating pulse envelope has been experimentally synthesized and used to resolve two consecutive temporal events, showing a temporal super-resolution measurement in comparison with a Gaussian signal.

In regard to the response of a physical system to a super-oscillating perturbation, some previous works have theoretically showed that the super-oscillatory characteristics of an electromagnetic field can be explored through the dynamics of a quantum system driven by such a field~\cite{Baranov2014,kempf2017driving}. In addition, it is argued that this particular behavior can be exploited, for instance, to induce atomic transitions by driving the system with an optical pulse, whose spectrum is below the transition line and to study ultrafast physical processes by means of low frequency signals.

Following these approaches, we present here a study of the interaction between an ultra-short pulse, as presented in Ref.~\cite{neyra2021tailoring,arXivsubfourier}, and a two-level quantum system (TLS). In the cited works, a simple way to experimentally obtain Gaussian beams and ultra-shorts laser pulses, that exhibit super-oscillatory characteristics, has been discussed and implemented. By using a modified Michelson interferometer, which adds a quadratic phase (chirp) in one of its arms, different pulses were synthesized. In the present letter we show that, for this kind of pulses, the phenomenon of super-oscillation can be extended as a global broadening of the effective bandwidth in the super-oscillatory region, and not only to the generation of a new frequency, higher than the highest Fourier component of the pulse, as is usually presented in the literature.

The state of a TLS is in general a superposition between the ground ($\ket{0}$) and the excited ($\ket{1}$) states,  $\ket{\psi(t)}=C_0(t)\ket{0}+C_1(t)e^{-i\omega_e t}\ket{1}$, where $\omega_e$ is the transition frequency, and $C_{0,1}(t)$ are the probability amplitudes to find the system in one of these eigenstates. The dynamic of this system, when it is driven by an external classical field, for instance the electric field of a laser pulse, can be described by the set of differential equations: 
\begin{eqnarray}\label{e1}
\frac{dC_0(t)}{dt}&=&i\Omega(t)e^{-i\omega_e t}C_1(t) \\  
\frac{dC_1(t)}{dt}&=&i\Omega^{*}(t)e^{i\omega_e t}C_0(t), \nonumber   
\end{eqnarray}
assuming that the relaxation times of the system are longer than the temporal width of the interacting pulse. 
The function $\Omega(t)$ is the well-known Rabi frequency, $\Omega(t)=\frac{X_{01}}{\hbar}E(t)$,
where $X_{01}$ represents the dipole matrix element between the states $\ket{0}$ and $\ket{1}$, and $E(t)$ is the time-dependent electric field,  that can be described as $E(t)=E_0\mathcal{E}(t)e^{i\omega_0 t}e^{i\phi}$, being $E_0$ the peak field amplitude, $\mathcal{E}(t)$ the field envelope, $\omega_0$ the carrier frequency and $\phi$ an arbitrary phase. 

\begin{figure*}[!ht]
\centering
\includegraphics[width=1.0\textwidth]{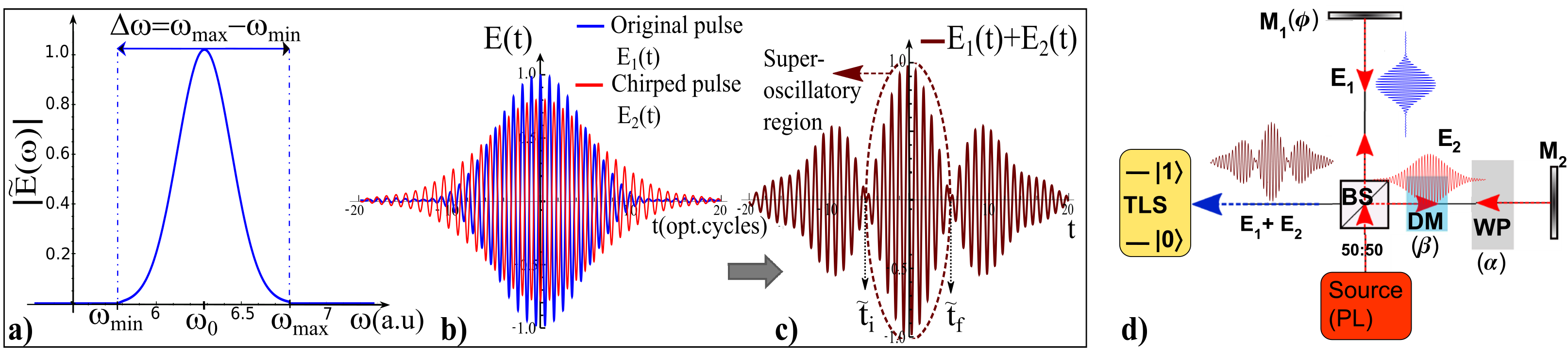}
\caption{a) FT amplitude corresponding to the pulses $E_1(t)$ and $E_2(t) $, both spectrally limited by $\Delta \omega =1$. b) Temporal evolution of $E_1(t)$ and $E_2(t)$. c) Pulse resulting from the interference between $E_1(t)$ and $E_2(t)$, for $\alpha =0.9 $ and $\beta = 10$. The super-oscillatory region, enclosed by a dashed line, is delimited by the roots of the field envelope $\mathcal{E}(t)$. d) Schematic of the setup to experimentally obtain the super-oscillatory pulse $E(t)=E_1(t)+E_2(t)$. Its shape is tuned by setting the values $\alpha$, $\beta$ and $\phi$ by means of a waveplate (WP), a dispersive media (DM) and a mirror $M_1$, respectively.} \label{profile}
\end{figure*}

We now consider a field $\tilde{E}(\omega)$ in the frequency domain, that can be synthesized in an analogous way to that followed in Ref.~\cite{neyra2021tailoring}. Schematically, the synthesis is performed by a Michelson-type interferometer, like the one sketched in Fig.~\ref{profile}(d). Here, the initial band-limited pulse, $\tilde{E}_1(\omega)$, that travels through the interferometer without being modified, is recombined with $\tilde{E}_2(\omega)$, which can be seen as the initial pulse modified by the chirp parameter ($\beta$), the amplitude ratio between the two arms ($\alpha$), and their relative phase ($\theta$). Thus, the resulting field is 
\begin{eqnarray}
 \tilde{E}(\omega)&=& E_0\left(e^{-10 (\frac{\omega - \omega_0}{\Delta\omega})^2}+\alpha e^{-10 (\frac{\omega - \omega_0}{\Delta\omega})^2}e^{i\beta(\omega - \omega_0)^2}e^{i\theta}\right)\nonumber\\
 &\times& \mathrm{rect}\left(\frac{\omega - \omega_0}{\Delta\omega}\right)=\tilde{E}_1(\omega)+\tilde{E}_2(\omega), \label{e2} 
\end{eqnarray}
where the envelope function $\mathrm{rect}(x)$ is the so-called rectangle function, which limits the frequency content of the pulse to the spectral range $\omega_{min}=\omega_0-\Delta\omega/2<\omega<\omega_0+\Delta\omega/2=\omega_{max}$ \cite{PhysRevA.103.053124}. Besides, the full
width at half maximum (FWHM) of the light intensity profile $|\tilde{E}(\omega)|^2$, is FWHM$_\omega=\sqrt{\frac{\ln(2)}{5}}\Delta\omega$.

\begin{figure}[h!]
\includegraphics{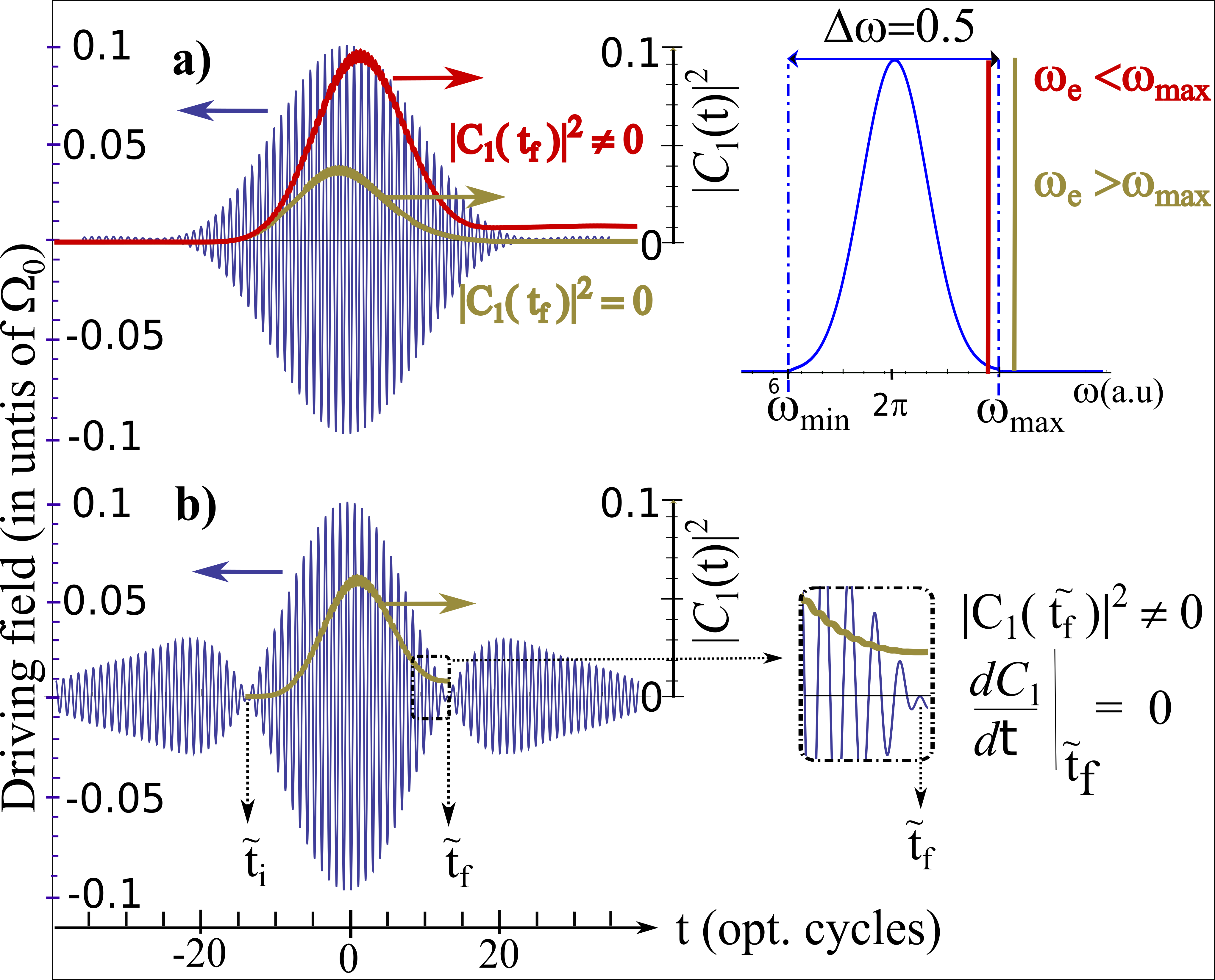}
\caption{Simulations of the time evolution of the excited state population of an ideal TLS. The system interacts with an external field with $\omega_0=2\pi$ and $\Delta \omega = 0.5$.  
In panel (a), the interaction occurs with the original pulse $E_1(t)$ ($\alpha=0$), when the transition frequency $\omega_e$ is below (red line) and above (gold line) $\omega_{max}$, so that, $\Delta=0.22$ and 0.30, respectively. In panel (b), the system is driven by a synthesized pulse with $\alpha = 0.6$, $\beta = 80$ and $\theta=0.1$, being $\omega_{max}<\omega_e$ ($\Delta=0.30$).} \label{fig2}
\end{figure}

\begin{figure*}
\centering
\includegraphics[width=1.0\textwidth]{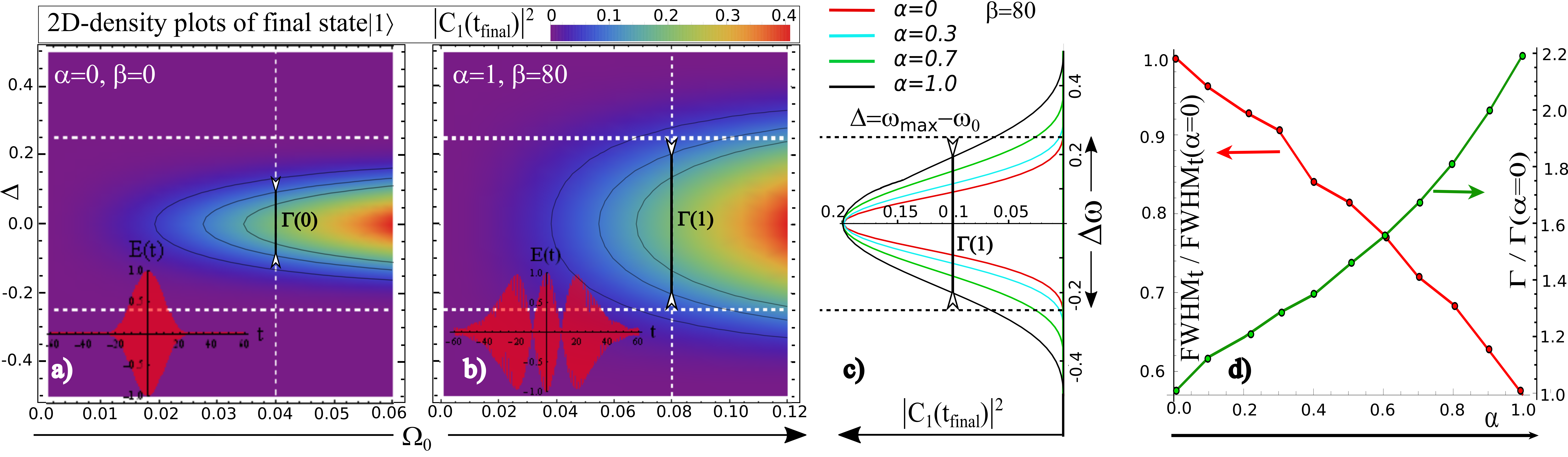}
\caption{2D-density plots of the final population of the excited state as a function of the electric field peak amplitude (in units of the strength of the Rabi frequency $\Omega_0$) and the detuning from resonance, $\Delta$. The TLS interacts with the primitive pulse $E_1(t)$ in panel (a), and with a super-oscillatory pulse in panel (b). The region between the horizontal dotted lines corresponds to transition frequencies within the spectral range of $E_1(t)$. c) Cross section of the 2D-density map displayed in panel (a) (red line) and in panel (b) (black line). Cross sections, corresponding to cuts of the 2D-density maps at $\alpha=0.3$ and $\alpha=0.7$, are also plotted. 
d) $\mathrm{FWHM}_t$ (red points) and temporally localized spectral FWHM, $\Gamma$ (green points). Each point corresponds to a different value of $\alpha$, and both $\mathrm{FWHM}_t$ and $\Gamma$ are measured relative to their values for $\alpha=0$. The values of $\Gamma$ are obtained from the corresponding cross section of $|C(\tilde{t}_f)|^2$.\label{fig3}} 
\end{figure*}

As a result of the boundary effects imposed by the rectangle function in Eq.~(\ref{e2}), the FT of $\tilde{E}(\omega)$ no longer corresponds to the interference between a Gaussian and a ``broadened" Gaussian pulse (chirped pulse). In fact, the FT of the function
$e^{-(\frac{\omega}{\Delta\omega_1})^2}\cross\mathrm{rect}(\frac{\omega}{\Delta\omega})$ is given by $E_{qG}(t)\propto e^{- \frac{1}{4}(t\Delta\omega_1)^2}[\ \mathrm{erfi}(\frac{\Delta\omega_1t}{2}-i\frac{\Delta\omega}{2\Delta\omega_1})
-\mathrm{erfi}(\frac{\Delta\omega_1t}{2}+i\frac{\Delta\omega}{2\Delta\omega_1})],$
where $\mathrm{erfi}(t)$ is the imaginary error function. The sum of the two $\mathrm{erfi}$ functions 
smoothly modifies the Gaussian pulse $e^{-\frac{1}{4}(t\Delta\omega_1)^2}$, if $\Delta\omega$ is large enough compared to $\Delta\omega_1$. In our case, this condition is fulfilled because $\Delta\omega>\Delta\omega_1=\Delta\omega/\sqrt{10}$, where the factor $\sqrt{10}$ comes from the argument of the Gaussian functions in Eq.~(\ref{e2}). Therefore, in the time domain, we can consider the pulse $E(t)$ (the FT of the field $\tilde{E}(\omega)$) as the interference of two ``quasi-Gaussian pulses", $E_1(t)$ and $E_2(t)$. The ratio between the temporal widths of the pulses $E_1(t)$ and $E_2(t)$ is, approximately, $(1+(\frac{\Delta\omega}{\sqrt{10}})^4\beta^2)^{1/2}$~\cite{neyra2021tailoring}.

In Figs.~\ref{profile}(a)-(c), we show a schematic representation of the initial band-limited pulse, 
the broadened pulse, and the synthesized super-oscillatory pulse, for the set of parameters $\Delta\omega=1$, $\beta=10$, $\alpha=0.9$ and $\theta=0$, respectively. The carrier frequency $\omega_0$ was fixed to $2\pi$.
Figure~\ref{profile}(a) shows the amplitude of the pulses spectral content, in the frequency domain, which is shared by both  $E_1(t)$ and $E_2(t)$. 
In Fig.~\ref{profile}(b), these pulses are represented in the time domain. Although both pulses have the same spectral content (in amplitude), the chirped pulse is temporarily broadened, relative to the primitive one. Finally, the Fig.~\ref{profile}(c) shows the synthesis obtained by the destructive interference between $E_1(t)$ and $E_2(t)$, giving origin to a super-oscillation region in the central part of the pulse, delimited by the times where the envelope of the wave becomes zero.

In general, the Eqs.~(\ref{e1}) are solved between an initial state at  $t_i=-\infty$ and a final state at $t_f=\infty$, two \textit{stationary points} in which the field does not interact with the system. Using first order perturbation theory (weak interacting fields), the coefficient $C_1(t \rightarrow \infty)$ 
results proportional to the FT of the envelope function $\tilde{\mathcal{E}}(\omega_e)$ (see for example Ref.~\cite{allen1987optical}), meaning that, after the pulse, the population of the excited state in an ideal TLS not vanishes only when the transition frequency $\omega_e$ is within the Fourier spectrum of the field. 
However, this is not the best approach to understand the behavior of the system under a super-oscillatory pulse, since super-oscillations appear in a finite temporal region. 
Instead of that,
we have evaluated the Eqs.~(\ref{e1}) between two other times, $\tilde{t}_i$ and $\tilde{t}_f$, where the field amplitude of our super-oscillatory pulse vanishes or, equivalently, $E(t)=0$, independently of the global phase $\phi$. These times, which  correspond to a pure destructive interference between the fields $E_1(t)$ and $E_2(t)$ (see Fig.~\ref{profile}(c)), are the ones that delimit the super-oscillatory region. Furthermore, it follows from Eqs.~(\ref{e1}) that $\left.\frac{dC_0(t)}{dt}\right|_{t=\tilde{t}_i}=\left.\frac{dC_1(t)}{dt}\right|_{t=\tilde{t}_f}=0$, implying that the coefficients $C_0(t)$ and $C_1(t)$ are constants around these times. Hence, these also constitute two stationary points where the Rabi oscillations are momentarily stopped. The subsequent revival of these oscillations, due to the fact that the field increases, for example after $\tilde{t}_i$, drives the system from a new initial state corresponding to $C_1(\tilde{t}_i)$. We can, thus, say that the evolution of the TLS in the super-oscillatory region is independent of the evolution between $t=-\infty$ and $t=\tilde{t}_i$.

In order to extract particular features of the synthesized super-oscillatory pulse, $E(t)=E_1(t)+E_2(t)$, we have numerically analyzed the solution of Eqs.~(\ref{e1}) by comparing the effects this particular pulse produces on the TLS and the ones driven by the original pulse, $E_1(t)$. As an example, in Fig.~\ref{fig2} it is shown the temporal evolution of the excited state population, $|C_1(t)|^2$, and the corresponding driving field. 
For all cases the system is initially in the ground state ($C_1=0$), and interacts with a relatively weak field. 
The values for the set of parameters, in units of $\omega_0=2\pi$, were chosen to be $\Delta\omega=0.5$ and $\Omega_0=0.1$, being $\Omega_0$ the strength of the Rabi frequency, which is proportional to the electric field amplitude. More specifically, if we work with a Ti:sapphire laser with a central period $T=2.7$fs ($\lambda\approx 800$nm), and in the case of Gaussian pulses where $\mathrm{FWHM}_t\times\mathrm{FWHM}_\omega=2\pi 0.441$, this would correspond to a pulse with a FWHM$_t\approx 15$ opt. cycles, that is a FWHM$_t\approx40$fs.

We start by considering, in Fig.~\ref{fig2}(a), the effect of $E_1(t)$ on the probability to find the system in the excited state, when the value of the detuning from resonance, defined by $\Delta=\omega_e-\omega_0$, is $0.22$ (red line). In this case, the interaction takes place in such a way that the transition frequency is within the spectral content of the field. The displayed result corresponds to the solution of Eqs.~(\ref{e1}) calculated between $t=t_i\rightarrow -\infty$ and $t=t_f\rightarrow \infty$. At $t=t_f$, it is observed that $|C_1|^2\neq0$ in agreement with the pulse area theorem~\cite{allen1987optical}.
Moreover, when the detuning is $\Delta=0.30$ (golden line), so the transition frequency is beyond the highest frequency of the pulse spectrum $\omega_{max}$, we obtain $|C_1(t_f)|^2=|C_1(t_i)|^2=0$, which indicates that the system remains in the ground state once the pulse ceases. 
Finally, Fig.~\ref{fig2}(b) shows the temporal evolution of $|C_1|^2$ for a detuning parameter value $\Delta=0.30$ as used before, but when the TLS is driven by a synthesized pulse that exhibits a super-oscillatory behavior. The parameter values used in the synthesis are $\beta=80$ and $\alpha=0.6$. 
As we previously discussed, the dynamic of the system in the super-oscillatory region will depend on the characteristics of the pulse in that temporal window, but not on the whole pulse. Hence, the solution is calculated between the two roots of $\mathcal{E}(t)$, $t=\tilde{t}_i$ and $t=\tilde{t}_f$, as indicated in the figure.
It can be seen that $|C_1(\tilde{t}_f)|^2\neq 0$ and $\left.\frac{dC_1(t)}{dt}\right|_{t=\tilde{t}_f}=0$.
In what follows, we will show that this non-zero population of the excited state at $t=\tilde{t}_f$ can be linked to a global broadening of the effective bandwidth in the time window where the super-oscillation occurs.

With the purpose of characterizing the effective local increase in the bandwidth, that we have called \textit{super-bandwidth}, we studied the dynamics of the TLS as a function of the detuning $\Delta$, and the peak amplitude of the electric field, $E_0$, measured in units of $\Omega_0$. In Fig.~\ref{fig3}, the 2D-density 
plots display the final value of the excited state population $|C_1(t)|^2$, that is reached by driving the system with the synthesized field. The calculations were performed, as previously described, between the roots of the field envelope, $\tilde{t}_i$ and $\tilde{t}_f$. In Fig.~\ref{fig3}(a), the set of the synthesis parameters is $\alpha=\beta=0$, which corresponds to the original pulse $E_1(t)$. We can see that, for weak fields, $|C_1(\tilde{t}_f)|^2$ does not vanish in a region delimited by $|\Delta|<0.25=\Delta\omega/2$ (horizontal dotted lines), so that the transition frequency of the TLS is within the spectral range of the original pulse, $\omega_{min}<\omega_e<\omega_{max}$.  
In Fig.~\ref{fig3}(b), the values of the synthesis parameters were set to $\alpha=1$, and $\beta=80$, resulting in a super-oscillatory pulse. In this case, the area where $|C_1(\tilde{t}_f)|^2\neq 0$ is broader than that in Fig.~\ref{fig3}(a) and extends beyond the limit $|\Delta|=0.25$, that is, $|\omega_e-\omega_0|>\Delta\omega/2$. 

The weak field approximation predicts a slight variation on the population of the excited state, during the time the pulse interacts with the system. Quantitatively, this means that the period of the Rabi oscillations, $T_0=2\pi/\Omega_0$, is appreciable longer than the FWHM$_t$. We have taken as a reference that such variation is at most 20\%.
This value is reached (vertical dotted lines) when $\Omega_0\approx0.04$ in the case of Fig.~\ref{fig3}(a), and $\Omega_0\approx 0.08$ in the case of Fig.~\ref{fig3}(b), that implies a Rabi period of $T_0=157$ opt. cycles (FWHM$_t=14.9\ll 157=T_0$), and $T_0=79$ opt. cycles (FWHM$_t=8.8\ll 79=T_0$), respectively.
In Fig.~\ref{fig3}(c), it can be seen a vertical cross section, for the value $\Omega_0$ leading to $|C_1(t)|^2\leq0.20$, of the 2D-density plots corresponding to a synthesized pulse with $\beta=80$ and different values of the parameter $\alpha$ (only the 2D-density plots for $\alpha=0$ and $\alpha=1$ are shown here). As a reference, for these value of $\beta$ and $\Delta\omega$, the chirped pulse is $\sqrt{5}\approx$ 2.2 times wider than the original pulse.

From each cross section, we can obtain the effective spectral full with at half maximum, $\Gamma$, corresponding to the super-oscillatory time window $\Delta t=\tilde{t}_f-\tilde{t}_i$. As expected, for $\alpha=0$ the bandwidth matches the value $\sqrt{\frac{5}{\ln(2)}}\Gamma=0.5=\Delta\omega$. 
For higher values of $\alpha$,
$|C_1(\tilde{t}_f)|^2\neq0$ beyond the spectral limit imposed by the FT, and the cross sections show new frequencies below $\omega_{min}$ and above $\omega_{max}$. Along with this, $\Gamma$ also increases. 
Finally, in Fig.~\ref{fig3} (d), we plotted both $\Gamma$ and FWHM$_t$ as a function of $\alpha$, until $\alpha=1$, when the super-oscillatory region has the same amplitude that the side-lobes (see inset in Fig.~\ref{fig3} (b)). 
It can be observed a monotonous decrease of $\mathrm{FWHM}_t$ with respect to its value at $\alpha =0$ (temporal super-resolution \cite{Boyko2005,Eliezer2017}), while $\Gamma$ steadily increases (super-bandwidth). 

In conclusion, by studying the evolution of a TLS in the interval between the zero crossings of the electric field envelope, we are able to neglect the effects of the interaction with the side lobes and individualized those effects purely originated in the super-oscillatory window. The simulations predict that, in this temporal region, there is a net final population of the excited level, even for a transition frequency beyond the Fourier spectral components of the pulse. We have shown that this behavior is due to the fact that the synthesized pulse 
has a temporally localized ``effective bandwidth", larger than its FT bandwidth, and not only to the appearance of a single frequency locally oscillating faster than its highest Fourier component, as is usually considered in the literature.      

It is worth mentioning that the pulse analyzed in this work shares similar characteristics, in the time domain, with that of Ref.~\cite{Boyko2005,Eliezer2017,eliezer2018experimental} which also exhibit a localized fast oscillation and a narrow temporal structure. 
Furthermore, from the synthesis method followed in Ref.~\cite{neyra2021tailoring,arXivsubfourier}, it is possible to obtain \textit{dichromatic pulses}, such as those used in Refs.~\cite{he2019coherently,koong2021coherent} for exciting a TLS with no spectral overlap with its optical transition.
In fact, from Eq.~(\ref{e2}), if $\alpha=1$ and imposing the interference condition $\theta=\pi$  when $\omega=\omega_0$, then $\tilde{E}(\omega_0)=0$, independently of $\beta$. 
Finally, we want to emphasize that the analysis carried out here is valid for pulses with an arbitrary FWHM$_t$, i.e.~in the multi-cycle regime. For few-cycles pulses, a deeper analysis is necessary, to take into account the effect of the carrier envelope phase \cite{PhysRevA.103.053124}.  
These facts are promising for future experimental implementations and can open new ways in coherent control, as well as in ultra-fast spectroscopy. We can mention that a preliminary analysis in high-harmonic generation in solids driven by the pulses studied in this work, shows novel and intriguing results. This ongoing research will be published elsewhere~\cite{Ciappina2021}.


\textbf{Funding} 
This work was partially supported by Agencia de Promoción Científica y Tecnológica under the project PICT 2016-4086. PIP 15-17 0435 (CONICET-Argentina) is acknowledged.

\textbf{Disclosures} The authors declare no conflicts of interest.


\bibliography{report} 
\bibliographystyle{spiebib} 
\end{document}